\documentclass[sigconf]{acmart}

\usepackage{hyperref}
\usepackage{float}
\usepackage{balance}
\usepackage{comment}
\usepackage{breakcites}
\hypersetup{breaklinks=true}

\def\BibTeX{{\rm B\kern-.05em{\sc i\kern-.025em b}\kern-.08emT\kern-.1667em\lower.7ex\hbox{E}\kern-.125emX}}

\usepackage[noline]{algorithm2e}
\SetKwInOut{Input}{input}
\SetKwInOut{Output}{output}
\newlength\inoutlen

\usepackage{graphicx}
\usepackage{subcaption}
\usepackage{epsf}
\newcommand{\epsin}[2]%
{\setlength{\epsfxsize}{#2\hsize}\centerline{\epsfbox{#1}}}
\renewcommand\footnotetextcopyrightpermission[1]{}

\usepackage{paralist}
\usepackage{tabularx}
\usepackage{flushend}
\usepackage{microtype}

\begin{document}
\fancyhead{}
\let\url\nolinkurl

\title{Detecting silent data corruptions in the wild}

\author{Harish Dattatraya Dixit}
\affiliation{%
\institution{Meta Platforms, Inc. }}
\email{hdd@fb.com}

\author{Laura Boyle}
\affiliation{%
\institution{Meta Platforms, Inc. }}
\email{lauraboyle@fb.com}

\author{Gautham Vunnam}
\affiliation{%
\institution{Meta Platforms, Inc. }}
\email{gauthamvunnam@fb.com}

\author{Sneha Pendharkar}
\affiliation{%
\institution{Meta Platforms, Inc. }}
\email{spendharkar@fb.com}

\author{Matt Beadon}
\affiliation{%
\institution{Meta Platforms, Inc. }}
\email{mbeadon@fb.com}

\author{Sriram Sankar}
\affiliation{%
\institution{Meta Platforms, Inc. }}
\email{sriramsankar@fb.com}

\renewcommand{\shortauthors}{Dixit et al.}

\begin{abstract}
Silent Errors within hardware devices occur when an internal defect manifests in a part of the circuit which does not have check logic to detect the incorrect circuit operation. The results of such a defect can range from flipping a single bit in a single data value, up to causing the software to execute the wrong instructions. Silent data corruptions (SDC) in hardware impact computational integrity for large-scale applications. Manifestations of silent errors are accelerated by datapath variations, temperature variance, and age, among other silicon factors. These errors do not leave any record or trace in system logs. As a result, silent errors stay undetected within workloads, and their effects can propagate across several services, causing problems to appear in systems far removed from the original defect. In this paper, we describe testing strategies to detect silent data corruptions within a large scale infrastructure. Given the challenging nature of the problem, we experimented with different methods for detection and mitigation. We compare and contrast two such approaches - 1. Fleetscanner (out-of-production testing) and 2. Ripple (in-production testing). We evaluate the infrastructure tradeoffs associated with the silicon testing funnel across 3+ years of production experience.
\end{abstract}


\maketitle
\renewcommand{\shortauthors}{Dixit \textit{et al.}}

\section{Introduction}
\label{s:introduction}

Meta Infrastructure serves numerous applications like Facebook, Whatsapp, Instagram, Messenger and Oculus workloads. All these applications expect computational integrity and reliability from the underlying infrastructure. Silent data corruptions challenge these fundamental assumptions and impact applications at scale. In our previous paper ~\citep{dixit2021silent}, we shared insights on the impact of silent data corruptions with a case study within the Spark workloads at Meta. In the shared example, a simple computation like $(1.1)^{53}$ resulted in the wrong answer (0 instead of 156.24), resulting in missing rows within the database, which subsequently led to data loss for the application. Within Meta infrastructure, we have observed hundreds of instances of unique silent data corruptions. Meta runs several detection and testing frameworks, and we prevent the impact to our applications before the corruption can propagate. We have employed these detection strategies since 2019 within our fleet. Within this paper, we provide insights into the different strategies which majorly fall into 2 buckets: 1. Out-of-production testing and 2. In-production testing. We summarize the tradeoffs and test metrics associated with different stages within the silicon lifecycle.

The paper is structured in the following way - Section ~\ref{s:related_work} provides insights on related work within this domain. Section ~\ref{s:silicon_testing_metrics} dives deep into the testing philosophies at different stages within the silicon lifecycle. Section ~\ref{s:motivation} provides the motivation for fleetwide testing. Section ~\ref{s:infra_testing} elaborates on the infrastructure testing strategies employed at Meta by exploring the in-production and out-of-production testing mechanisms. Section ~\ref{s:results} provides insights into the results associated with the different strategies and evaluates tradeoffs associated with them. Section ~\ref{s:conclusion} concludes the paper.

\section{Related Work}
\label{s:related_work}

There has been recent interest in the area of silent data corruptions ~\citep{dixit2021silent}, ~\citep{hochschild2021cores}, ~\citep{DBLP:journals/corr/abs-2110-11519}.
Prior studies ~\citep{Baumann}, ~\citep{SEP-uC}, ~\citep{IBM} within this domain focused on the soft errors induced due to cosmic rays. Fault injection studies ~\citep{ARM-R5-VA}, ~\citep{Bitflipinjection}, ~ \citep{elliott2013quantifying} focused on fault modeling using soft error occurrence rates which were modeled at one fault in a million silicon devices. Meta published one of the first studies on large scale impact of silent errors ~\citep{dixit2021silent}, and showed that the SDC occurrence rate of one in thousand silicon devices is reflective of fundamental silicon challenges, and not limited to particle effects or cosmic rays. Google also published their observations ~\citep{hochschild2021cores}, where mercurial cores were identified to disobey the fundamental rules of computation and produce erroneous results. This is an industry wide problem. At OCP 2021 ~\cite{OCP}, a panel of experts ~\cite{OCPPanel} from both industry and academia within the silent error domain gathered to discuss the strategies for the domain moving forward. More research and articles ~\cite{SELSE2021}, ~\cite{ITC2021},  ~\cite{nextplatformarticle}, ~\cite{NYTarticle}, ~\cite{Intelrelease} establish the importance of this domain. The research focus for industry and academia has strongly been on identifying strategies and mitigating silent data corruptions not only in CPUs but also in all silicon devices.

\section{Silicon testing funnel}
\label{s:silicon_testing_metrics}
\begin{figure*}[ht!]
\centering
\includegraphics[width=\textwidth, keepaspectratio=true, scale = 0.55]{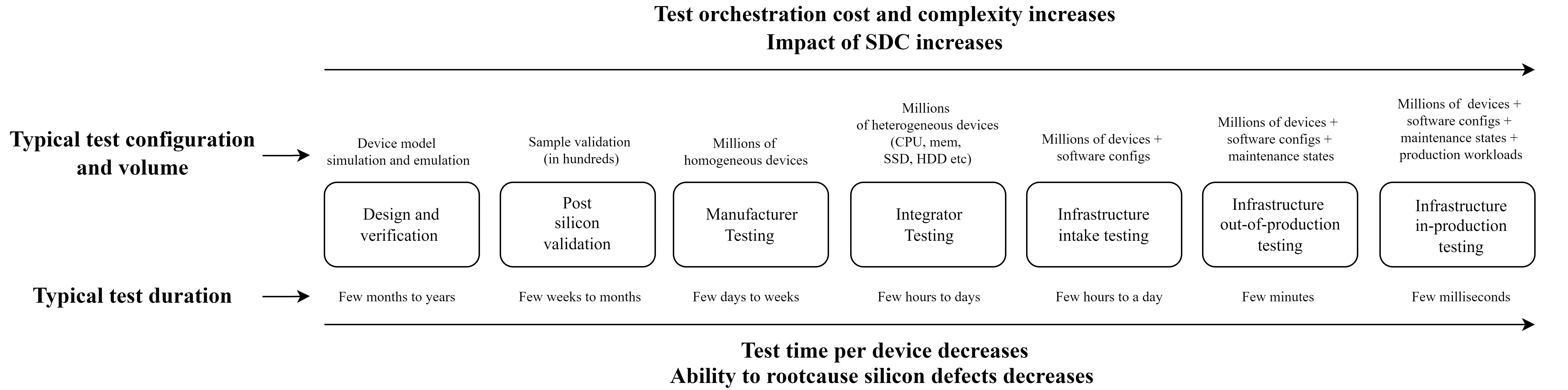}
\caption{Silicon testing funnel}
\label{f:silicon testing funnel}
\end{figure*}

Before a silicon device reaches the Meta infrastructure fleet, the silicon device goes through different stages as part of the silicon development process. In this section, we will not go into elaborate details regarding the silicon development process. We are comparing the testing strategies employed at different stages to understand the cost associated with testing at fleetwide scale, and why that can be challenging. The \textbf{silicon testing funnel} in figure ~\ref{f:silicon testing funnel} provides a high level comparison for different stages within this section. Following subsections provide primitive descriptions of the different stages. It is to be noted that each stage is a dedicated research topic on its own, and the testing model varies for hyperscalars versus enterprise scale companies. In this paper, we focus on three important parameters: testing volume, test time, and the impact of a fault at that stage.

\subsection{Design and verification}
\label{ss:design_verif}
For any silicon device, once the architectural requirements are finalized, the silicon design and development process is initiated. Testing is usually limited to a few design models of the device, and simulations and emulations are used to test different features. The device is tested regularly with implementation of novel features. Test iterations are implemented on a daily basis. The cost of testing is low relative to the other stages, and the testing is repeated using different silicon variation models. Design iteration at this stage is faster than any other stage in the process. Faults can be identified based on internal states that are not visible in later stages of the development cycle. The test cost increases slowly with placement of standard cells for ensuring that the device meets the frequency and clock requirements, and also with the addition of different physical characteristics associated with the materials as part of the physical design of the device. There is plenty of available industry research on testing optimizations within these stages ~\citep{lam2005hardware}, ~\citep{668981}, ~\citep{larsson2002integrated}, ~\citep{3141}, ~\citep{hunt1989microprocessor}. The testing process here lasts usually for many months to a couple of years depending on the chip and the development stages employed.

\subsection{Post silicon validation}
\label{ss:post_si}
At this stage, numerous device samples are available for validation. Using the test modes available within the design of the device, the design is validated for different features. The number of device variations has grown from models in the previous stage to actual physical devices exhibiting manufacturing variance. Significant fabrication costs have been incurred before obtaining the samples, and a device fault at this stage has a higher impact since it typically results in a re-spin for the device. Additionally, there is a larger test cost associated with precise and expensive instrumentation for multiple devices under test. At the end of this validation phase, the silicon device can be considered as approved for mass production. The testing process here typically lasts for a few weeks to a few months.

\subsection{Manufacturer testing}
\label{ss:manufacturer}
At mass production, every device is subjected to automated test patterns using advanced fixtures. Based on the results of the testing patterns, the devices are binned into different performance groups to account for manufacturing variations. As millions of devices are tested and binned, time allocated for testing has a direct impact on manufacturing throughput. The testing volume has increased from a few devices in the previous step to millions of devices, and test cost scales per device. Faults are expensive at this stage, as they typically result in respin or remanufacturing of the device.

\subsection{Integrator testing}
\label{ss:integrator_testing}
After the manufacturing and testing steps, the devices are shipped to an end customer. A large scale infrastructure operator typically utilizes an integrator to coordinate the process of rack design, rack integration and server installation. The integrator facility typically conducts testing for multiple sets of racks at once. The complexity of testing has now increased from one device type to multiple types of devices working together in cohesion. The test cost increases from a single device to testing for multiple configurations and combinations of multiple devices. An integrator typically tests the racks for a few days to a week. Any faults require reassembly of racks and reintegration. 

\subsection{Infrastructure intake testing}
\label{ss:infra_intake_test}
As part of the rack intake process, infrastructure teams typically conduct an intake test where the entire rack received from the integrator is wired together with datacenter networks within the designated locations. Subsequently, test applications are executed on the device before executing actual production workloads. In testing terms, this is referred to as infrastructure burn-in testing. Tests are executed for a few hours to a couple of days. There are hundreds of racks containing a large number of complex devices that are now paired with complex software application tools and operating systems. The testing complexity has increased significantly relative to previous test iterations. A fault is challenging to diagnose due to the larger source of fault domain.

\subsection{Infrastructure fleet testing}
\label{ss:infra_fleet_test}
Historically, the testing practices concluded at infrastructure burn-in testing. The device is expected to work for the rest of its lifecycle, and any faults if observed would be captured using system health metrics and reliability-availability-serviceability features built into devices, which allow for collecting system health signals.

However, with silent data corruptions, there is no symptom or signal that indicates there is a fault with a device. Hence without running dedicated test patterns to detect and triage silent data corruptions, it is almost impossible to protect an infrastructure application from corruption. As a result, it has become imperative to test periodically within the fleet using different strategies. At this point within the lifecycle, the device is already part of a rack and serving production workloads. The testing cost is high relative to other stages, as it requires complex orchestration and scheduling while ensuring that the workloads are drained and undrained effectively. Tests are designed to run in complex multi-configuration, multi-workload environments. Any time spent in creating test environments and running the tests is time taken away from server running production workloads.

A fault within a production fleet is expensive to triage and root-cause as the fault domains have evolved to be more complex with ever changing software and hardware configurations. As a result, advanced strategies are required to detect silent data corruptions with expensive infrastructure tradeoffs.

\section{Why is this a hard problem ?}
\label{s:motivation}

With millions of devices, within a large scale infrastructure, there is a probability of error propagation to the applications. With an occurrence rate of one fault within a thousand devices, silent data corruptions have the ability to impact numerous applications. Until the application exhibits noticeable difference at higher level metrics, the corruption continues to propagate and produce erroneous computations. This scale of fault propagation presents a significant challenge to a reliable infrastructure. We have observed that faults can be due to a variety of sources or accelerants. We categorize these into four major sections. We turn to periodic testing with dynamic control of tests to triage corruptions and protect applications. These observations are based on testing and aggregating samples for $\approx$3 years within Meta infrastructure. We are using an example product computation of \textit{3 times 5} to demonstrate our observations:

\begin{itemize}
\item\textbf{Data randomization:}
We observe that the corruptions are data dependent by nature. For example, we observe numerous instances where the majority of the computations would be fine within a corrupt CPU but a smaller subset would always produce faulty computations due to certain bit pattern representation. For example, we may observe that 3 times 5 is 15, but 3 times 4 is evaluated to 10, and thus until and unless 3 times 4 is verified specifically, we cannot confirm computation accuracy within the device for that specific computation. This leads to a fairly large state space for testing.

\item\textbf{Electrical variations:}
In a large scale infrastructure, with varying nature of workloads and scheduling algorithms, the devices undergo a variety of operating frequency (f), voltage (V) and current (I) fluctuations. We observe that changing operating voltages, frequency and current associated with the device can lead to acceleration of occurrence of erroneous results on faulty devices. While the result would be accurate with one particular set of f, V and I, the result may not hold true for all the possible operating points. This leads to a multi-variate state space. For example, we may observe that 3 times 5 is 15 in some operating conditions, but repeating the same calculation may not always result in 15 under all operating conditions.

\begin{figure*}[h!]
\centering
\includegraphics[width=\textwidth, keepaspectratio=true, scale = 0.34]{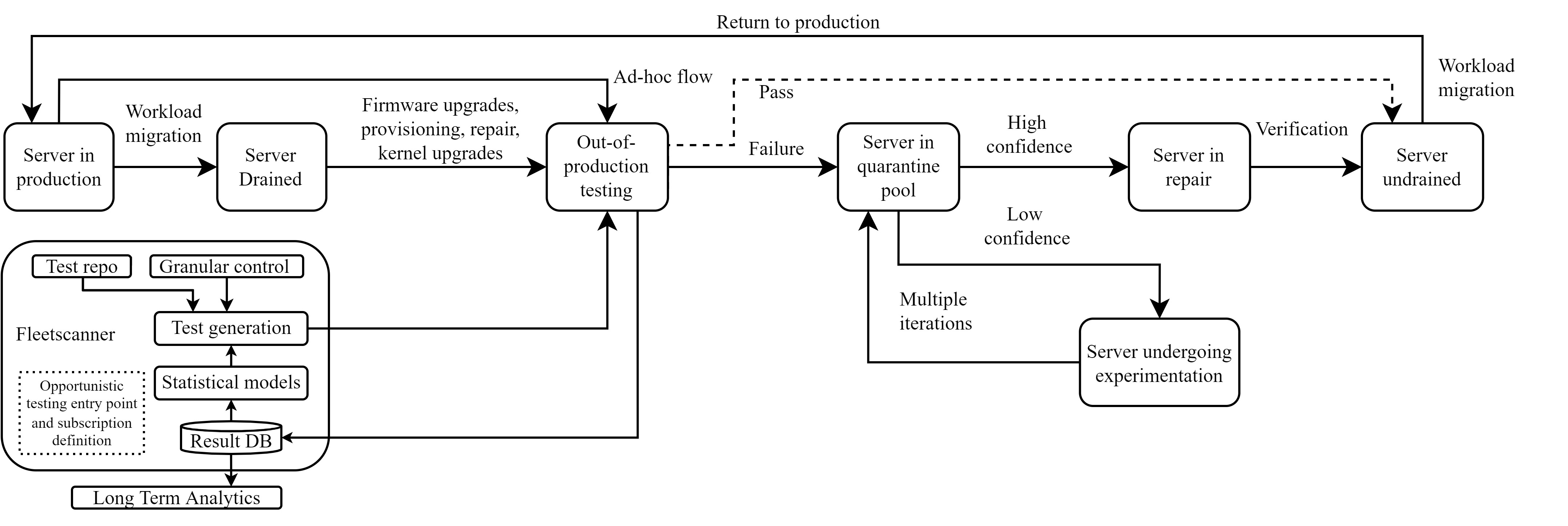}
\caption{Out-of-production testing}
\label{f:out_of_production_testing}
\end{figure*}

\item\textbf{Environmental variations:}
We observe that variations in location dependent parameters also accelerate occurrence of silent data corruptions. It is well documented that temperature  ~\citep{humenay2006impact}, ~\citep{wolpert2012temperature},
~\citep{zorn2014temperature}, ~\citep{schroder2003negative}, and humidity ~\citep{schroder2015semiconductor}, ~\citep{cimmino2020high}, ~\citep{1479958} have a direct impact on the voltage and frequency parameters associated with the device due to device physics. In a large-scale datacenter, while the temperature and humidity variations are controlled to be minimal, there can be occurrences of hot-spots within specific server locations due to the nature of repeated workloads on that server and neighboring servers. Also the seasonal trends associated with a datacenter location can create hotspots across data halls within a datacenter. For example, we may observe 3 times 5 is 15 in datacenter A, but repeated computations can result in 3 times 5 computing to 12 in datacenter B.

\item\textbf{Lifecycle variations:}
We observe that silicon continually changes in performance and reliability with time. This has been well documented in bathtub curve failure modeling across the literature ~\citep{klutke2003critical}, ~\citep{roesch2012using}, ~\citep{pecht2017guidebook}. However, with silent data corruptions we observe that certain failures can manifest earlier than the traditional bathtub curve predictions based on device usage. As a result, a computation producing a correct result today provides no guarantee that the computation will produce a correct result tomorrow. In one specific experiment, we repeated the exact same computation sequence on the device once every day for a period of 6 months and the device failed after 6 months indicating degradation with time for that computation. In essence, a computation like 3 times 5 equals 15 can provide a correct result today but tomorrow may result in 3 times 5 being evaluated to an incorrect value.

\end{itemize}

As a result of all four observations, we conclude that the only way to measurably protect the fleet against silent data corruptions is to repeatedly test the infrastructure with ever improving test routines and advanced test pattern generation. By building engineering capability in finding hidden patterns across hundreds of failures, and feeding the insights into optimizations for test runtimes, testing policies and architectures, the fleet resiliency can be improved. Sharing these insights with vendors, industry and academia on a periodic basis also enables the collective research growth within this domain.

\section{Infrastructure Testing}
\label{s:infra_testing}
As part of Meta infrastructure, we have implemented 2 broad categories of testing at fleet scale. When a fleet is made up of millions of machines spread across multiple regions and fault domains, it is important that testing is efficient and tactical. The 2 broad categories of testing are:

\begin{itemize}
\item Out-of-production testing.
\item In-production testing.
\end{itemize}

\subsection{Out-of-production testing}
\label{ss:oop_opportunistic}

Out-of-production testing refers to the ability to subject machines to known patterns of inputs, and comparison of its expected outputs with known reference values across millions of different execution paths. Tests are executed across different temperatures, voltages, machine types, regions etc. while the machine is idle and not executing production workloads. 

The test patterns are generated based on our production experience and understanding of silicon architectures as well as obtained from silicon vendors. The instructions are carefully crafted in sequences to match known defects or target a variety of defect families using numerous state search policies within the testing state space.

Typically in a large scale infrastructure, there are always sets of machines going through maintenance. Before any of these maintenance are started, the workload is safely migrated off the machine, typically referred to as a draining phase. Post a successful drain phase, we observe one or many of the following maintenance:

\begin{itemize}
\item\textbf{Firmware upgrades:}
There are numerous devices within a given server and there may be new firmware available on at least one component. These component firmware upgrades are required to keep the fleet up to date for fixing firmware bugs as well as security vulnerabilities.
\item\textbf{Kernel upgrades:}
Similar to component level upgrades, the kernel on a particular server is upgraded at a regular cadence, and these provide numerous application and security updates for the entire fleet.
\item\textbf{Provisioning:}
While the above two mechanisms refer to the process of upgrading a server. Provisioning refers to the process of preparing the server for workloads with installation of operating systems, drivers and application-specific recipes. There could also be instances of reprovisioning where-in within a dynamic fleet a server is moved from one type of workload to another.
\item\textbf{Repair:}
Each server that encounters a known fault or triggers a match to a failing signature ends up in a repair queue. Within the repair queue, based on the diagnoses associated with the device, a soft repair (without replacing hardware components) is conducted or a component swap is executed. This enables faulty servers to return back to production.
\end{itemize}

Any machine exiting the maintenance phase is then undrained to make the machine available to production workloads. With these maintenances already available within the fleet, we at Meta developed and integrated a tool called \textit{Fleetscanner}. Fleetscanner opportunistically identifies machines entering and exiting maintenance states and schedules the machines to undergo silent data corruption testing. The architecture for fleetscanner and its integration at a very high level is represented in Figure ~\ref{f:out_of_production_testing}. In all the cases, based on the time available and the type of machine identified, fleetscanner runs optimized versions of tests and provides a snapshot for the device’s response to sensitive architectural codepaths, and verifies the computations to be accurate. A number of machine specific parameters are captured at this instant to enable understanding the conditions that result in device failures. Any machine identified to fail for silent data corruption routines are routed to the quarantine pool for further investigation and test refinements.

\begin{figure*}[tb]
\centering
\includegraphics[width=\textwidth, keepaspectratio=true, scale = 0.24]{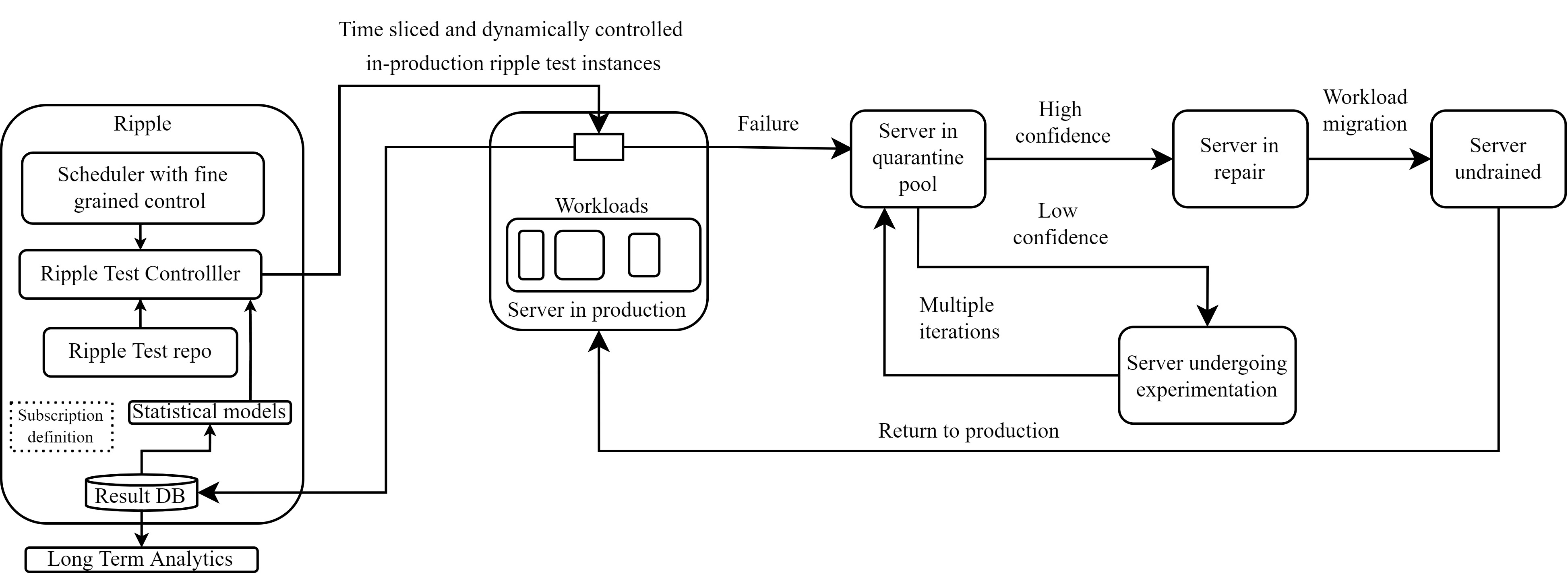}
\caption{In-production testing}
\label{f:in_production_testing}
\end{figure*}

The four out-of-production workflows are independent complex systems with orchestration across millions of machines, and fleetscanner enables a seamless methodology to orchestrate silent data corruption tests within a large fleet by integrating with all the workflows. It is extremely important to minimize the time spent in drain and undrain phases and piggyback on existing maintenance. It is also important to minimize disruption to existing workflows with significant time overheads and orchestration complexities. This allows the testing cost to be noticeable yet minimal per machine while providing reasonable protection against application corruptions.

\subsection{In-production testing}
\label{ss:ip_prod}
While out-of-production testing allows for testing opportunistically when machines transition across states, there are many instances within our fleet where a novel signature identified must be immediately scaled to the entire fleet. Waiting for out-of-production scanning opportunities and subsequently ramping up fleetwide coverage is slow. While fleetscanner has its own benefits in implementing longer running tests, with test runtime in minutes, we observe a requirement for an alternate light-weight method to test within the fleet while the machines are running production workloads. This is tricky to achieve without a granular understanding of the workload and modulation of testing routines with the workloads.

At Meta, we have implemented a testing methodology called \textit{Ripple} which co-locates with the workload, and executes test instructions for millisecond level intervals. The architecture for ripple testing is described in figure ~\ref{f:in_production_testing}. Test sequences used in out-of-production testing are modified specifically to be conducive to run through Ripple. Typically intrusive tests are used as part of infrastructure burn-in testing; changing them to run in ripple mode requires fine-tuning of tests along with test coverage tradeoff decisions. The test orchestration is implemented with extreme care as any variation within the test could immediately affect the application. This test is \textit{live} within the entire fleet and provides granular control on test subsets, cores to test, type of workloads to co-locate with as well as in scaling the test up and down to multiple sets of cores based on the workload.

\begin{figure}[hbpt]
\centering
\includegraphics[keepaspectratio=true, scale = 0.14]{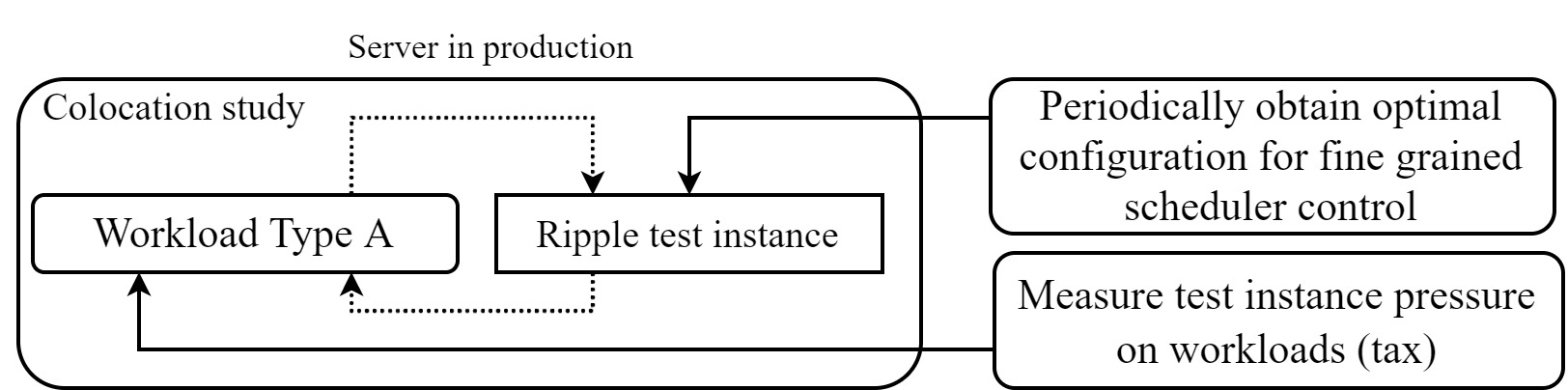}
\caption{Shadow testing}
\label{f:shadow_testing}
\end{figure}

\subsubsection{Shadow testing:}
\label{sss:shadow_testing}
We implemented and fully rolled out ripple across multiple sets of workloads. We have also crafted the \textit{ripple} test architecture to be able to have safeguards to prevent fleetwide fallout in case of a test defect. We implemented shadow testing by running a wide variety of workloads with A/B testing for different instruction sequences with different seasonality and across different workloads. A major challenge in shadow testing is enabling colocation. Based on the scaling of the workload, the testing mechanism had to descale. For each type of workload at Meta, we identified an evaluation process for the scaling factor. Based on instrumentation, we established the footprint tax associated with the test. Each workload type’s colocation study provides the tax, and the goal for the tool is to minimize its tax below a certain threshold. A simplified version of the architecture is referenced in Figure ~\ref{f:shadow_testing}. With repeated sets of experimentation, we established control structures and safeguards for enabling different options for different workloads, and then scaled the solution to the entire fleet.

\subsubsection{Always-on in production:}
\label{sss:always_on}
This mechanism is \textit{always on}. Only the scale at which it is operating is dynamically controlled through configurations within the fleet. This methodology is powerful in finding defects which require thousands of iterations of the same data inputs, as well as in identifying devices undergoing degradation. A novel signature identified within the fleet for a device could be scaled to the entire fleet with satisfiable test randomization and defect pattern matching within a couple of weeks. This methodology is also uniquely effective in identifying silicon transition defectsl. In the comparison of results below, we share the statistical value of this method in identifying silent data corruptions for a large subset of defects for one CPU defect family.

\subsubsection{Recommendations to the industry:}
\label{sss:recos_industry}
We have recommended this mechanism of testing as an important evolution for silent data corruption tools from large scale internal studies. Based on our findings and data-sharing practices, our vendors have enabled modes within their tests that make them suitable for ripple test. Vendors have published white-papers ~\cite{Intelrelease} around a testing methodology called \textit{trickle testing} which derives from the in-production testing flow and its fleetwide success within Meta. We would like to thank our industry partners in taking lessons from our fleetwide studies and making them available to industry and academia.

\begin{table*}[!ht]
\centering
\begin{tabularx}{\textwidth}{|| X | X | X ||}
\hline
\textbf{Metric} & \textbf{Fleetscanner} & \textbf{Ripple}\\
\hline\hline
\textbf{Total tests executed} & $\approx$68 million (lifetime) & $\approx$2.5 billion (per month) \\
\textbf{Testing time} & $\approx$4 billion fleet seconds (lifetime) & $\approx$100 million fleet seconds (per month) \\
\textbf{Performance aware} & No & Yes\\
\textbf{Unique SDC coverage} & 23 percent & 7 percent \\
\textbf{Time to equivalent SDC coverage} & $\approx$6 months (70 percent) & $\approx$15 days (70 percent) \\
\hline
\end{tabularx}
\caption{Comparison of Fleetscanner and Ripple}
\label{f:comparison_table}
\end{table*}

\section{Results}
\label{s:results}

We are sharing results below from around 3 years of data aggregation regarding the effectiveness of these two different testing strategies. The result dataset here in comparison is for a large subset of defects for one CPU defect family, with tests being executed on a significantly large percentage of the fleet. Equivalent coverage within this section refers to the ability to detect the same set of failures through different methods of testing.

\subsection{Out-of-production testing (Fleetscanner)}
Using fleetscanner as part of the fleetwide out-of-production detection across millions of machines, we have obtained a total of around 68 million unique test iterations within the lifetime of the tool. Test runtimes vary based on type of maintenance available and the type of integration sequence in place. In total, we have tested for around 4 billion test seconds. The tests are inherently intrusive in nature and hence conducted out-of-production. We observe that fleetscanner provides 93 percent coverage among all detected silent data corruptions for defect family under study. Fleetscanner also achieves 23 percent unique coverage which is not reachable by ripple. Based on different cadences of maintenance, we observe that some machines may undergo significantly more testing than others. However, fleetscanner achieves approximately full fleetwide coverage within five months to six months based on average deployments and maintenance.

From a test cost perspective, this is expensive. The fleet spends a significant compute time executing tests. However, it is our observation that this is an important cost with increasing sightings of silent data corruptions.

\subsection{In-production testing (Ripple)}
In comparison to the out-of-production testing, the ripple test framework provides its own set of unique coverage metrics. Since ripple is always-on, we are able to achieve around 2.5 billion unique test instances any given month because of the non-intrusive nature of the tests and granular control and co-location. Test runtimes vary based on workload intensity and the subscription configurations. However, given that each test is limited to hundreds of milliseconds at best, we obtain a total test runtime of around 100 million fleet seconds every month.

Ripple testing offers a unique coverage of 7 percent among the set of all detectable machines. We observe that this coverage is impossible to achieve with fleetscanner due to the inherent nature of testing and the underlying silicon defects. To elaborate, certain failures are detected via ripple due to frequent transitions of test instructions along with workloads, and are not detected with continuous long running tests. While 7 percent coverage is unique to ripple, it can detect 70 percent within the 93 percent coverage that fleetscanner provides within 15 days. While ripple can achieve this coverage within 15 days, fleetscanner requires around 5 to 6 months. This scaling effect makes ripple a powerful framework within a large fleet.



\subsection{Comparison}
A comparison of the numbers presented in the above 2 sections is provided in table ~\ref{f:comparison_table}. From the table, we observe that for defect family under study within this paper, 70 percent of the common coverage detection could be completed within 15 days using ripple. Fleetscanner ramps up to the remaining 23 percent of the coverage over 6 months. A unique 7 percent coverage is through repeated ripple instances within the fleet. Ripple provides a total coverage of 77 percent with significantly lower total test runtimes than fleetscanner. There are benefits to both models of testing. We also consistently revisit and evaluate these coverage metrics to inform and update our fleetwide testing strategies around test vectors, test cadences and test runtimes. We observe that with different types of defects, the coverage split varies.

Historically, each CPU only went through a few hours of testing as part of infrastructure burn-in tests. Further testing was typically conducted via sampling. We observe that novel detection approaches are required for application health and fleet resiliency. In this paper, we demonstrate the ability to test at scale and get through billions of fleet seconds of testing every month across a large fleet consistently. These novel techniques enable us to detect silent data corruptions and mitigate them at scale.

\section{Conclusions}
\label{s:conclusion}
Detecting silent data corruption is a challenging problem for large-scale infrastructures. Applications show significant sensitivity to these problems and can be exposed to such corruptions for months without accelerated detection mechanisms. It can also result in data loss and require months to debug and resolve software level residue of silent corruptions. This research shows novel techniques resulting from years of experience observing silent corruptions and in categorizing their occurrence patterns and faster time to detection. Impact of silent data corruption can have a cascading effect on applications and we have to address this as a critical problem. As a result, detecting these at scale as quickly as possible is important towards enabling a safer and reliable fleet.

\pagebreak

\textbf{Acknowledgement}
The authors would like to thank Aslan Bakirov, Melita Mihaljevic, Thiara Ortiz, Tejasvi Chakravarthy, Manish Modi, Vijay Rao, T.S. Khurana, Nishant Yadav, Aravind Anbudurai and other infrastructure engineers for their inputs in the implementation of solutions and valuable technical suggestions. The authors would also like to thank Bill Holland, Chris BeSerra, Marty Humphrey, Fred Lin and Daniel Moore for their thoughtful comments, feedback and efforts towards improving this paper. 

\bibliographystyle{ACM-Reference-Format}
\bibliography{selse2022.bib}


\begin{thebibliography}{31}


\ifx \showCODEN    \undefined \def \showCODEN     #1{\unskip}     \fi
\ifx \showDOI      \undefined \def \showDOI       #1{#1}\fi
\ifx \showISBNx    \undefined \def \showISBNx     #1{\unskip}     \fi
\ifx \showISBNxiii \undefined \def \showISBNxiii  #1{\unskip}     \fi
\ifx \showISSN     \undefined \def \showISSN      #1{\unskip}     \fi
\ifx \showLCCN     \undefined \def \showLCCN      #1{\unskip}     \fi
\ifx \shownote     \undefined \def \shownote      #1{#1}          \fi
\ifx \showarticletitle \undefined \def \showarticletitle #1{#1}   \fi
\ifx \showURL      \undefined \def \showURL       {\relax}        \fi
\providecommand\bibfield[2]{#2}
\providecommand\bibinfo[2]{#2}
\providecommand\natexlab[1]{#1}
\providecommand\showeprint[2][]{arXiv:#2}

\bibitem[\protect\citeauthoryear{??}{ITC}{2021}]%
        {ITC2021}
 \bibinfo{year}{2021}\natexlab{}.
\newblock \bibinfo{booktitle}{\emph{2021 Program – International Test
  Conference.}}
\newblock
\urldef\tempurl%
\url{http://www.itctestweek.org/wp-content/uploads/2021/10/2021-Final-Program.pdf}
\showURL{%
\tempurl}


\bibitem[\protect\citeauthoryear{??}{SEL}{2021}]%
        {SELSE2021}
 \bibinfo{year}{2021}\natexlab{}.
\newblock \bibinfo{booktitle}{\emph{2021 Program – Silicon Errors in Logic
  – System Effects.}}
\newblock
\urldef\tempurl%
\url{https://selse.org/previous-workshops/2021-archive/2021-program/}
\showURL{%
\tempurl}


\bibitem[\protect\citeauthoryear{??}{OCP}{2021a}]%
        {OCP}
 \bibinfo{year}{2021}\natexlab{a}.
\newblock \bibinfo{booktitle}{\emph{Open Compute Project. 2021}}.
\newblock
\urldef\tempurl%
\url{https://www.opencompute.org/summit/global-summit}
\showURL{%
\tempurl}


\bibitem[\protect\citeauthoryear{??}{OCP}{2021b}]%
        {OCPPanel}
 \bibinfo{year}{2021}\natexlab{b}.
\newblock \bibinfo{booktitle}{\emph{PANEL OCP HW Operation at Scale Reliability
  to Address Silent Data Corruptions.}}
\newblock
\urldef\tempurl%
\url{https://www.youtube.com/watch?v=3yhg4Gt8M\_E}
\showURL{%
\tempurl}


\bibitem[\protect\citeauthoryear{Abadir, Ferguson, and Kirkland}{Abadir
  et~al\mbox{.}}{1988}]%
        {3141}
\bibfield{author}{\bibinfo{person}{M.S. Abadir}, \bibinfo{person}{J. Ferguson},
  {and} \bibinfo{person}{T.E. Kirkland}.} \bibinfo{year}{1988}\natexlab{}.
\newblock \showarticletitle{Logic design verification via test generation}.
\newblock \bibinfo{journal}{\emph{IEEE Transactions on Computer-Aided Design of
  Integrated Circuits and Systems}} \bibinfo{volume}{7}, \bibinfo{number}{1}
  (\bibinfo{year}{1988}), \bibinfo{pages}{138--148}.
\newblock
\urldef\tempurl%
\url{https://doi.org/10.1109/43.3141}
\showDOI{\tempurl}


\bibitem[\protect\citeauthoryear{{Baumann}}{{Baumann}}{2005}]%
        {Baumann}
\bibfield{author}{\bibinfo{person}{R.~C. {Baumann}}.}
  \bibinfo{year}{2005}\natexlab{}.
\newblock \showarticletitle{Radiation-induced soft errors in advanced
  semiconductor technologies}.
\newblock \bibinfo{journal}{\emph{IEEE Transactions on Device and Materials
  Reliability}} \bibinfo{volume}{5}, \bibinfo{number}{3}
  (\bibinfo{year}{2005}), \bibinfo{pages}{305--316}.
\newblock
\urldef\tempurl%
\url{https://doi.org/10.1109/TDMR.2005.853449}
\showDOI{\tempurl}


\bibitem[\protect\citeauthoryear{{Bossen}}{{Bossen}}{2002}]%
        {IBM}
\bibfield{author}{\bibinfo{person}{D. {Bossen}}.}
  \bibinfo{year}{2002}\natexlab{}.
\newblock \showarticletitle{CMOS Soft Errors and Server Design - IRPS. Tutorial
  Notes - Reliability Fundamentals}.
\newblock  (\bibinfo{year}{2002}).
\newblock


\bibitem[\protect\citeauthoryear{{Cardarilli}, {Kaddour}, {Leandri}, {Ottavi},
  {Pontarelli}, and {Velazco}}{{Cardarilli} et~al\mbox{.}}{2002}]%
        {Bitflipinjection}
\bibfield{author}{\bibinfo{person}{G.~C. {Cardarilli}}, \bibinfo{person}{F.
  {Kaddour}}, \bibinfo{person}{A. {Leandri}}, \bibinfo{person}{M. {Ottavi}},
  \bibinfo{person}{S. {Pontarelli}}, {and} \bibinfo{person}{R. {Velazco}}.}
  \bibinfo{year}{2002}\natexlab{}.
\newblock \showarticletitle{Bit flip injection in processor-based
  architectures: a case study}. In \bibinfo{booktitle}{\emph{Proceedings of the
  Eighth IEEE International On-Line Testing Workshop (IOLTW 2002)}}.
  \bibinfo{pages}{117--127}.
\newblock
\urldef\tempurl%
\url{https://doi.org/10.1109/OLT.2002.1030194}
\showDOI{\tempurl}


\bibitem[\protect\citeauthoryear{Cimmino and Ferrero}{Cimmino and
  Ferrero}{2020}]%
        {cimmino2020high}
\bibfield{author}{\bibinfo{person}{Davide Cimmino} {and}
  \bibinfo{person}{Sergio Ferrero}.} \bibinfo{year}{2020}\natexlab{}.
\newblock \showarticletitle{High-voltage temperature humidity bias test
  (HV-THB): Overview of current test methodologies and reliability
  performances}.
\newblock \bibinfo{journal}{\emph{Electronics}} \bibinfo{volume}{9},
  \bibinfo{number}{11} (\bibinfo{year}{2020}), \bibinfo{pages}{1884}.
\newblock


\bibitem[\protect\citeauthoryear{Dixit, Pendharkar, Beadon, Mason,
  Chakravarthy, Muthiah, and Sankar}{Dixit et~al\mbox{.}}{2021}]%
        {dixit2021silent}
\bibfield{author}{\bibinfo{person}{Harish~Dattatraya Dixit},
  \bibinfo{person}{Sneha Pendharkar}, \bibinfo{person}{Matt Beadon},
  \bibinfo{person}{Chris Mason}, \bibinfo{person}{Tejasvi Chakravarthy},
  \bibinfo{person}{Bharath Muthiah}, {and} \bibinfo{person}{Sriram Sankar}.}
  \bibinfo{year}{2021}\natexlab{}.
\newblock \showarticletitle{Silent Data Corruptions at Scale}.
\newblock \bibinfo{journal}{\emph{arXiv preprint arXiv:2102.11245}}
  (\bibinfo{year}{2021}).
\newblock


\bibitem[\protect\citeauthoryear{Elliott, Mueller, Stoyanov, and
  Webster}{Elliott et~al\mbox{.}}{2013}]%
        {elliott2013quantifying}
\bibfield{author}{\bibinfo{person}{James Elliott}, \bibinfo{person}{Frank
  Mueller}, \bibinfo{person}{Frank Stoyanov}, {and} \bibinfo{person}{Clayton
  Webster}.} \bibinfo{year}{2013}\natexlab{}.
\newblock \bibinfo{booktitle}{\emph{Quantifying the impact of single bit flips
  on floating point arithmetic}}.
\newblock \bibinfo{type}{{T}echnical {R}eport}. \bibinfo{institution}{North
  Carolina State University. Dept. of Computer Science}.
\newblock


\bibitem[\protect\citeauthoryear{Gronowski, Bowhill, Preston, Gowan, and
  Allmon}{Gronowski et~al\mbox{.}}{1998}]%
        {668981}
\bibfield{author}{\bibinfo{person}{P.E. Gronowski}, \bibinfo{person}{W.J.
  Bowhill}, \bibinfo{person}{R.P. Preston}, \bibinfo{person}{M.K. Gowan}, {and}
  \bibinfo{person}{R.L. Allmon}.} \bibinfo{year}{1998}\natexlab{}.
\newblock \showarticletitle{High-performance microprocessor design}.
\newblock \bibinfo{journal}{\emph{IEEE Journal of Solid-State Circuits}}
  \bibinfo{volume}{33}, \bibinfo{number}{5} (\bibinfo{year}{1998}),
  \bibinfo{pages}{676--686}.
\newblock
\urldef\tempurl%
\url{https://doi.org/10.1109/4.668981}
\showDOI{\tempurl}


\bibitem[\protect\citeauthoryear{Hemsoth}{Hemsoth}{2021}]%
        {nextplatformarticle}
\bibfield{author}{\bibinfo{person}{Nicole. Hemsoth}.}
  \bibinfo{year}{2021}\natexlab{}.
\newblock \bibinfo{booktitle}{\emph{How Facebook Architects Around Silent Data
  Corruption}}.
\newblock
\urldef\tempurl%
\url{https://www.nextplatform.com/2021/03/01/facebook-architects-around-silent-data-corruption/}
\showURL{%
\tempurl}


\bibitem[\protect\citeauthoryear{Hochschild, Turner, Mogul, Govindaraju,
  Ranganathan, Culler, and Vahdat}{Hochschild et~al\mbox{.}}{2021}]%
        {hochschild2021cores}
\bibfield{author}{\bibinfo{person}{Peter~H Hochschild}, \bibinfo{person}{Paul
  Turner}, \bibinfo{person}{Jeffrey~C Mogul}, \bibinfo{person}{Rama
  Govindaraju}, \bibinfo{person}{Parthasarathy Ranganathan},
  \bibinfo{person}{David~E Culler}, {and} \bibinfo{person}{Amin Vahdat}.}
  \bibinfo{year}{2021}\natexlab{}.
\newblock \showarticletitle{Cores that don't count}. In
  \bibinfo{booktitle}{\emph{Proceedings of the Workshop on Hot Topics in
  Operating Systems}}. \bibinfo{pages}{9--16}.
\newblock


\bibitem[\protect\citeauthoryear{Humenay, Tarjan, and Skadron}{Humenay
  et~al\mbox{.}}{2006}]%
        {humenay2006impact}
\bibfield{author}{\bibinfo{person}{Eric Humenay}, \bibinfo{person}{David
  Tarjan}, {and} \bibinfo{person}{Kevin Skadron}.}
  \bibinfo{year}{2006}\natexlab{}.
\newblock \bibinfo{booktitle}{\emph{Impact of parameter variations on
  multi-core chips}}.
\newblock \bibinfo{type}{{T}echnical {R}eport}. \bibinfo{institution}{VIRGINIA
  UNIV CHARLOTTESVILLE DEPT OF COMPUTER SCIENCE}.
\newblock


\bibitem[\protect\citeauthoryear{Hunt}{Hunt}{1989}]%
        {hunt1989microprocessor}
\bibfield{author}{\bibinfo{person}{Warren~A Hunt}.}
  \bibinfo{year}{1989}\natexlab{}.
\newblock \showarticletitle{Microprocessor design verification}.
\newblock \bibinfo{journal}{\emph{Journal of automated reasoning}}
  \bibinfo{volume}{5}, \bibinfo{number}{4} (\bibinfo{year}{1989}),
  \bibinfo{pages}{429--460}.
\newblock


\bibitem[\protect\citeauthoryear{{Iturbe}, {Venu}, and {Ozer}}{{Iturbe}
  et~al\mbox{.}}{2016}]%
        {ARM-R5-VA}
\bibfield{author}{\bibinfo{person}{X. {Iturbe}}, \bibinfo{person}{B. {Venu}},
  {and} \bibinfo{person}{E. {Ozer}}.} \bibinfo{year}{2016}\natexlab{}.
\newblock \showarticletitle{Soft error vulnerability assessment of the
  real-time safety-related ARM Cortex-R5 CPU}. In
  \bibinfo{booktitle}{\emph{2016 IEEE International Symposium on Defect and
  Fault Tolerance in VLSI and Nanotechnology Systems (DFT)}}.
  \bibinfo{pages}{91--96}.
\newblock
\urldef\tempurl%
\url{https://doi.org/10.1109/DFT.2016.7684076}
\showDOI{\tempurl}


\bibitem[\protect\citeauthoryear{Klutke, Kiessler, and Wortman}{Klutke
  et~al\mbox{.}}{2003}]%
        {klutke2003critical}
\bibfield{author}{\bibinfo{person}{Georgia-Ann Klutke},
  \bibinfo{person}{Peter~C Kiessler}, {and} \bibinfo{person}{Martin~A
  Wortman}.} \bibinfo{year}{2003}\natexlab{}.
\newblock \showarticletitle{A critical look at the bathtub curve}.
\newblock \bibinfo{journal}{\emph{IEEE Transactions on reliability}}
  \bibinfo{volume}{52}, \bibinfo{number}{1} (\bibinfo{year}{2003}),
  \bibinfo{pages}{125--129}.
\newblock


\bibitem[\protect\citeauthoryear{Lam}{Lam}{2005}]%
        {lam2005hardware}
\bibfield{author}{\bibinfo{person}{William~KC Lam}.}
  \bibinfo{year}{2005}\natexlab{}.
\newblock \bibinfo{booktitle}{\emph{Hardware design verification: simulation
  and formal method-based approaches}}.
\newblock \bibinfo{publisher}{Prentice Hall Professional Technical Reference}.
\newblock


\bibitem[\protect\citeauthoryear{Larsson, Peng, and Chakrabarty}{Larsson
  et~al\mbox{.}}{2002}]%
        {larsson2002integrated}
\bibfield{author}{\bibinfo{person}{Erik Larsson}, \bibinfo{person}{Zebo Peng},
  {and} \bibinfo{person}{Krishnendu Chakrabarty}.}
  \bibinfo{year}{2002}\natexlab{}.
\newblock \showarticletitle{An integrated framework for the design and
  optimization of SOC test solutions}.
\newblock In \bibinfo{booktitle}{\emph{SOC (System-on-a-Chip) Testing for Plug
  and Play Test Automation}}. \bibinfo{publisher}{Springer},
  \bibinfo{pages}{21--36}.
\newblock


\bibitem[\protect\citeauthoryear{Markoff}{Markoff}{2022}]%
        {NYTarticle}
\bibfield{author}{\bibinfo{person}{John Markoff}.}
  \bibinfo{year}{2022}\natexlab{}.
\newblock \bibinfo{booktitle}{\emph{Tiny Chips, Big Headaches}}.
\newblock
\urldef\tempurl%
\url{https://www.nytimes.com/2022/02/07/technology/computer-chips-errors.html}
\showURL{%
\tempurl}


\bibitem[\protect\citeauthoryear{{Mukherjee}, {Emer}, and
  {Reinhardt}}{{Mukherjee} et~al\mbox{.}}{2005}]%
        {SEP-uC}
\bibfield{author}{\bibinfo{person}{S.~S. {Mukherjee}}, \bibinfo{person}{J.
  {Emer}}, {and} \bibinfo{person}{S.~K. {Reinhardt}}.}
  \bibinfo{year}{2005}\natexlab{}.
\newblock \showarticletitle{The soft error problem: an architectural
  perspective}. In \bibinfo{booktitle}{\emph{11th International Symposium on
  High-Performance Computer Architecture}}. \bibinfo{pages}{243--247}.
\newblock
\urldef\tempurl%
\url{https://doi.org/10.1109/HPCA.2005.37}
\showDOI{\tempurl}


\bibitem[\protect\citeauthoryear{Pecht, Radojcic, and Rao}{Pecht
  et~al\mbox{.}}{2017}]%
        {pecht2017guidebook}
\bibfield{author}{\bibinfo{person}{Michael~G Pecht}, \bibinfo{person}{Riko
  Radojcic}, {and} \bibinfo{person}{Gopal Rao}.}
  \bibinfo{year}{2017}\natexlab{}.
\newblock \bibinfo{booktitle}{\emph{Guidebook for managing silicon chip
  reliability}}.
\newblock \bibinfo{publisher}{CRC press}.
\newblock


\bibitem[\protect\citeauthoryear{Roesch}{Roesch}{2012}]%
        {roesch2012using}
\bibfield{author}{\bibinfo{person}{William~J Roesch}.}
  \bibinfo{year}{2012}\natexlab{}.
\newblock \showarticletitle{Using a new bathtub curve to correlate quality and
  reliability}.
\newblock \bibinfo{journal}{\emph{Microelectronics Reliability}}
  \bibinfo{volume}{52}, \bibinfo{number}{12} (\bibinfo{year}{2012}),
  \bibinfo{pages}{2864--2869}.
\newblock


\bibitem[\protect\citeauthoryear{Sbar and Kozakiewicz}{Sbar and
  Kozakiewicz}{1979}]%
        {1479958}
\bibfield{author}{\bibinfo{person}{N.L. Sbar} {and} \bibinfo{person}{R.P.
  Kozakiewicz}.} \bibinfo{year}{1979}\natexlab{}.
\newblock \showarticletitle{New acceleration factors for temperature, humidity,
  bias testing}.
\newblock \bibinfo{journal}{\emph{IEEE Transactions on Electron Devices}}
  \bibinfo{volume}{26}, \bibinfo{number}{1} (\bibinfo{year}{1979}),
  \bibinfo{pages}{56--71}.
\newblock
\urldef\tempurl%
\url{https://doi.org/10.1109/T-ED.1979.19380}
\showDOI{\tempurl}


\bibitem[\protect\citeauthoryear{Schroder}{Schroder}{2015}]%
        {schroder2015semiconductor}
\bibfield{author}{\bibinfo{person}{Dieter~K Schroder}.}
  \bibinfo{year}{2015}\natexlab{}.
\newblock \bibinfo{booktitle}{\emph{Semiconductor material and device
  characterization}}.
\newblock \bibinfo{publisher}{John Wiley \& Sons}.
\newblock


\bibitem[\protect\citeauthoryear{Schroder and Babcock}{Schroder and
  Babcock}{2003}]%
        {schroder2003negative}
\bibfield{author}{\bibinfo{person}{Dieter~K Schroder} {and}
  \bibinfo{person}{Jeff~A Babcock}.} \bibinfo{year}{2003}\natexlab{}.
\newblock \showarticletitle{Negative bias temperature instability: Road to
  cross in deep submicron silicon semiconductor manufacturing}.
\newblock \bibinfo{journal}{\emph{Journal of applied Physics}}
  \bibinfo{volume}{94}, \bibinfo{number}{1} (\bibinfo{year}{2003}),
  \bibinfo{pages}{1--18}.
\newblock


\bibitem[\protect\citeauthoryear{Serebryany, Lifantsev, Shtoyk, Kwan, and
  Hochschild}{Serebryany et~al\mbox{.}}{2021}]%
        {DBLP:journals/corr/abs-2110-11519}
\bibfield{author}{\bibinfo{person}{Kostya Serebryany}, \bibinfo{person}{Maxim
  Lifantsev}, \bibinfo{person}{Konstantin Shtoyk}, \bibinfo{person}{Doug Kwan},
  {and} \bibinfo{person}{Peter Hochschild}.} \bibinfo{year}{2021}\natexlab{}.
\newblock \showarticletitle{SiliFuzz: Fuzzing CPUs by proxy}.
\newblock \bibinfo{journal}{\emph{CoRR}}  \bibinfo{volume}{abs/2110.11519}
  (\bibinfo{year}{2021}).
\newblock
\showeprint[arXiv]{2110.11519}
\urldef\tempurl%
\url{https://arxiv.org/abs/2110.11519}
\showURL{%
\tempurl}


\bibitem[\protect\citeauthoryear{Van De~Ven}{Van De~Ven}{2021}]%
        {Intelrelease}
\bibfield{author}{\bibinfo{person}{Arjan. Van De~Ven}.}
  \bibinfo{year}{2021}\natexlab{}.
\newblock \bibinfo{booktitle}{\emph{Intel Data Center Diagnostic Tool Now
  Available for Deployment}}.
\newblock
\urldef\tempurl%
\url{https://www.intel.com/content/www/us/en/developer/articles/tool/intel-data-center-diagnostic-tool-now-available.html}
\showURL{%
\tempurl}


\bibitem[\protect\citeauthoryear{Wolpert and Ampadu}{Wolpert and
  Ampadu}{2012}]%
        {wolpert2012temperature}
\bibfield{author}{\bibinfo{person}{David Wolpert} {and} \bibinfo{person}{Paul
  Ampadu}.} \bibinfo{year}{2012}\natexlab{}.
\newblock \showarticletitle{Temperature effects in semiconductors}.
\newblock In \bibinfo{booktitle}{\emph{Managing temperature effects in
  nanoscale adaptive systems}}. \bibinfo{publisher}{Springer},
  \bibinfo{pages}{15--33}.
\newblock


\bibitem[\protect\citeauthoryear{Zorn and Kaminski}{Zorn and Kaminski}{2014}]%
        {zorn2014temperature}
\bibfield{author}{\bibinfo{person}{Christian Zorn} {and} \bibinfo{person}{Nando
  Kaminski}.} \bibinfo{year}{2014}\natexlab{}.
\newblock \showarticletitle{Temperature humidity bias (THB) testing on IGBT
  modules at high bias levels}. In \bibinfo{booktitle}{\emph{CIPS 2014; 8th
  International Conference on Integrated Power Electronics Systems}}. VDE,
  \bibinfo{pages}{1--7}.
\newblock


\end{thebibliography}

\end{document}